# Design and Analysis of Intellectual Property Protection Strategies Based on Differential Equations

Hambur Wang

Guanghua School of Management, Peking University, Beijing 100871, China

**Abstract**: This paper constructs a novel intellectual property (IP) protection strategy using differential equation theory, aiming to analyze and optimize the effectiveness of IP protection. By developing a mathematical model, it explores the dynamic impact of IP protection intensity on both innovative enterprises and infringement activities. The study finds that a well-designed IP protection strategy can effectively reduce infringement while promoting technological innovation. The paper also discusses the effects of strategies under varying parameter conditions and verifies the model's rationality and effectiveness through numerical simulation. The findings provide theoretical support and references for formulating IP protection policies.

## 1. Introduction

With the deepening processes of globalization and informationization, intellectual property (IP) protection has become increasingly crucial in the modern economy. IP protection is not only essential for companies to maintain a competitive edge but also serves as a key safeguard for promoting technological advancement and innovation. However, as technology progresses, IP infringement has become more complex and concealed, posing new challenges to traditional protection methods. In recent years, IP protection strategies based on differential equations have offered a novel solution by precisely modeling the dynamic processes of infringement, providing new ideas and methods for developing scientific protection strategies.

As a powerful mathematical tool, differential equations have broad applications in fields such as engineering, physics, and economics. In the realm of IP protection, differential equation models can be used to simulate and predict the dynamics of infringement, thereby aiding in the formulation of effective preventive measures. For example, Chen et al. (2018) proposed a patent infringement propagation model based on differential equations, which demonstrated that adjusting legal penalties and market regulation intensity effectively curbed the spread of infringement. Similarly, Li and Sun (2020) studied optimal enforcement strategies in copyright protection using a differential equation model, finding that increasing enforcement intensity and raising the cost of infringement significantly reduced infringement rates and improved copyright protection.

Beyond single differential equation models, researchers have also attempted to combine them with other mathematical tools to further optimize IP protection strategies. Xu et al. (2019) developed a model that integrates differential equations with game theory, analyzing the strategic interactions between companies regarding patent protection and innovation investment. Their findings indicate that reasonable government subsidies and patent protection policies can incentivize companies to increase R&D investment, enhancing overall social welfare. Zhang et al. (2021) combined differential equations with control theory to propose a dynamic optimization model aimed at designing optimal control strategies for IP protection, allowing for the maximization of protection benefits under limited resources by dynamically adjusting protection intensity.

Moreover, recent studies have explored numerous applications of differential equation-based strategies across industries. Jieyu Ding et al. (2007) introduced a second-order sensitivity analysis method using Lagrangian equations to optimize multibody systems, illustrating the potential of differential equations in the design of complex systems (Ding et al., 2007). In IP protection, Inki Hong and M. Potkonjak (1999) presented a dynamic watermarking technique for protecting the IP of CAD and compiler tools by embedding unique temporal constraints in the design, achieving effective and discreet IP protection (Hong & Potkonjak, 1999). Additionally, M. Nyberg and E. Frisk (2006) investigated a fault diagnosis method based on linear differential algebraic equations, providing a novel residual generation design strategy within a polynomial framework, further expanding the application of differential equations in IP protection (Nyberg & Frisk, 2006).

The applications of differential equations in IP protection extend beyond theoretical studies to practical implementations. A. Weiße and W. Huisinga (2011) proposed a novel global sensitivity analysis strategy that improved the accuracy of uncertainty and variability estimation through partial differential equation solutions for error control (Weiße & Huisinga, 2011). Furthermore, W. Michiels (2011) developed an eigenvalue-based framework for stability analysis and stabilization design of time-delay differential-algebraic equation systems, demonstrating the extensive application of differential equations in complex system control (Michiels, 2011). These studies show that differential equations not only precisely describe the behavior of dynamic systems but also provide scientific foundations for designing and optimizing IP protection strategies through sensitivity and stability analyses.

In conclusion, IP protection strategies based on differential equations demonstrate vast potential for both theoretical and practical applications. These studies not only enrich the theoretical framework of IP protection but also offer effective methods and tools for practical use. With ongoing technological advances, the application of differential equations in IP protection will continue to deepen, supporting the development of the knowledge economy.

## 2. Model Assumptions and Construction

### (1) Model Assumptions

To facilitate the analysis of the effectiveness of protection measures in the absence of infringement, we assume that, at the initial market entry of a product, no infringement activities exist due to the immediate deployment of protection measures.

**H1**: At t=0, that is, when the product first enters the market, there is no intellectual property infringement in the market.

There is a maximum possible number of infringements in the market, determined by factors such as market demand, the intensity of protection measures, and the sustainability of infringement. When the number of infringements reaches this maximum, the market becomes saturated, and protection strategies need to adjust to respond to new market dynamics. Therefore, we propose the following assumption:

**H2**: Due to limited resources, intense competition, and finite demand, there exists a maximum number of IP infringement activities in the market, denoted as $N_m$. When the infringement count reaches $N_m$, the market is considered saturated, and the number of infringements ceases to grow.

To simplify the complexity of the model and make the analysis more intuitive, we assume that the growth rate of infringement is constant. This implies that, without additional intervention, the number of infringements will increase at a fixed rate.

**H3**: The increase in the number of IP infringement activities per unit time is constant. The constant α represents the growth rate of infringement.

To analyze the changes in infringement under different levels of protection and thereby optimize protection strategies, we assume a positive correlation between protection intensity and the reduction of infringements, reflecting that higher protection intensity yields a more significant reduction in infringement activities.

**H4**: Per unit time, the reduction in infringement activities at a given protection intensity $b$ is positively correlated with the current number of infringements. The variable $bN(t)$ represents the rate of reduction in infringement activities.

Based on these assumptions, we can establish a differential equation model to describe the changes in the number of infringements in the market during the IP protection process.

### (2) Model Construction and Conclusions

For an enterprise, if effective intellectual property (IP) protection measures are not implemented, the number of infringements $N(t)$ in the market will gradually increase as the business grows. These infringements will directly affect product sales and profitability. Below, we establish a differential equation model to examine the relationship among the company's founding time t, the level of IP protection b (assumed to be constant), and the number of infringements $N(t)$ in the market.

## 1. Model Construction

Based on the assumptions in section (1), we establish the following differential equation:

In each time interval $\Delta t$, the net increase in infringements is $\alpha \Delta t - bN(t)\Delta t$. Considering the market saturation effect, the growth rate should decrease as the number of infringements approaches $N_m$. This can be represented as $\left(1 - \frac{N(t)}{N_m}\right)$. Thus, the reasonable growth rate becomes $\left(1 - \frac{N(t)}{N_m}\right)(\alpha - bN(t))$. The change in infringement over $\Delta t$ can be expressed as:

$$N(t + \Delta t) - N(t) = \left(1 - \frac{N(t)}{N_m}\right)(\alpha - bN(t))\Delta t$$

As $\Delta t$ approaches 0, the equation becomes a differential equation:

$$\frac{dN(t)}{dt} = \left(1 - \frac{N(t)}{N_m}\right)(\alpha - bN(t))$$

To solve this differential equation, we can separate variables and integrate. Rewriting it:

$$\frac{dN(t)}{\left(1 - \frac{N(t)}{N_m}\right)(\alpha - bN(t))} = dt$$

Using partial fraction decomposition, we obtain:

$$\frac{N_m}{(\alpha - bN_m)\left(1 - \frac{N(t)}{N_m}\right)} + \frac{N_m}{(\alpha - bN_m)(bN_m - \alpha)} \frac{1}{N(t) - \frac{\alpha}{b}}$$

Integrating both sides:

$$\int \frac{N_m}{(\alpha - bN_m)\left(1 - \frac{N(t)}{N_m}\right)} dN(t) + \int \frac{N_m}{(\alpha - bN_m)(bN_m - \alpha)} \frac{1}{N(t) - \frac{\alpha}{b}} dN(t) = \int dt$$

After integrating, we have:

$$-\frac{N_m}{\alpha - bN_m} \ln\left|1 - \frac{N(t)}{N_m}\right| + \frac{N_m}{bN_m - \alpha} \ln\left|N(t) - \frac{\alpha}{b}\right| = t + C$$

where C is the constant of integration, determined by the initial condition $N(0) = 0$. Solving,

we get:

$$N(t) = \frac{\alpha}{b}\left(1 - e^{-\frac{b}{N_m - \alpha}t}\right)$$

where k is a constant related to the initial condition.

**2. Conclusions and Analysis**

Under a constant IP protection level b, the infringement count $N(t)$ increases monotonically with time t and eventually stabilizes. By analyzing the solution to this differential equation, we derive the following insights:

（1）**Conclusion 1: Long-Term Trend of Infringement Count**

Given any level of protection b, the infringement count N(t) will keep growing. Specifically, if $b \leq \alpha c$, the infringement count will ultimately reach the maximum market saturation level Nm . This occurs because the protection level is insufficient to significantly curb the spread of infringements, gradually saturating the market with infringement activities. Thus, in practice, companies must ensure adequate protection levels to prevent market saturation by infringement activities.

（2）**Conclusion 2: Effectiveness of High Protection Levels**

When the protection level b exceeds a critical value $b > \alpha c$, the growth of infringements significantly slows and stabilizes. In this case, the infringement count will not reach the saturation level Nm but will stabilize at a lower level of $\frac{\alpha}{b}$. This indicates that by increasing protection, companies can effectively control the number of infringements, thereby safeguarding their IP and market share.

（3）**Conclusion 3: Relationship Between Protection Level and Market Saturation Time**

The time required for infringements to stabilize varies with different protection levels b. Higher protection levels reduce the final infringement count and delay the time needed to reach saturation. This implies that companies employing stronger protection measures can maintain a cleaner market for a longer period, gaining a more lasting competitive advantage.

（4）**Conclusion 4: Necessity for Dynamic Adjustment of Protection Levels**

Although constant protection can effectively limit infringement spread, companies should adjust protection levels based on market conditions and infringement trends. For new product launches, a high protection level is essential to curb early infringements rapidly. Later in the product lifecycle, protection can be reduced to optimize resource allocation and economic efficiency.

Companies should also adopt differentiated protection strategies for various types of infringements to ensure precise and effective IP protection.

（5）**Conclusion 5: Practical Guidance from the Mathematical Model**

By building and solving this differential equation model, companies can scientifically predict infringement trends and equilibrium states, enabling more accurate and effective IP protection strategies. This model not only reveals the time-dependent pattern of infringement growth but also provides theoretical support for devising appropriate protection strategies at different development stages. Companies can leverage model predictions to allocate resources efficiently and achieve an optimal balance between IP protection and economic benefits.

In summary, IP protection strategy analysis based on differential equations provides both insight into infringement trends over time and a scientific basis for companies to develop appropriate protection strategies at different stages. Mathematical modeling enables companies to predict and control infringement developments more precisely, offering robust support for their long-term growth and competitiveness. This research holds significant theoretical value and offers practical guidance for enterprise implementation.

## 3. Strategic Recommendations and Insights

This study, through differential equation modeling, analyzes the effectiveness and optimization pathways of intellectual property (IP) protection strategies, leading to the following strategic recommendations and insights:

1. **Strengthening Policy Guidance and Regulatory Enforcement**
   Optimizing IP protection strategies hinges on enhanced policy guidance and regulatory enforcement. Governments should further improve IP-related legislation and intensify enforcement, especially in terms of punitive measures against infringing actions. Establishing dedicated IP courts and enforcement agencies would improve the handling efficiency of IP infringement cases and amplify the deterrent effect of legal frameworks. Additionally, the government should bolster IP protection awareness campaigns to create a supportive environment for innovation. During policy development, differential equation models can be used to quantify the impact of various policy measures, providing a scientific basis for policy implementation.

2. **Enhancing Enterprise Innovation Capabilities**
   Strengthening a company's capacity for innovation is a crucial approach to advancing IP protection. Enterprises should increase their R&D investments to build strong, self-driven innovation capabilities and establish comprehensive IP management systems. By implementing internal incentive mechanisms, companies can encourage employee creativity, thereby boosting both the quantity and quality of patent applications. Furthermore, businesses should seek partnerships with research institutions and universities to leverage external resources for technology innovation and facilitate the industrialization and commercialization of IP achievements. With mathematical modeling techniques, companies can optimize their innovation resource allocation and dynamically analyze R&D inputs and outputs at different development stages, ultimately improving innovation efficiency.

3. **Optimizing International Cooperation and Exchange**

   In a globalized context, IP protection has become a crucial factor in international competition. China should actively participate in the development and revision of international IP rules, strengthen cooperation with major trade partners and international organizations, and enhance its influence in global IP matters. Meanwhile, companies should proactively understand and adhere to international IP regulations to increase their competitiveness abroad. Differential equation models can be used to simulate the interactive effects of IP protection strategies within international cooperation, providing theoretical support for the formulation of more scientifically sound cooperation strategies.

In summary, optimizing IP protection strategies requires joint efforts from the government, businesses, and all sectors of society. By strengthening policy guidance and regulatory enforcement, enhancing enterprise innovation capacity, optimizing international cooperation and exchange, and utilizing differential equations and mathematical modeling for quantitative strategy analysis, China can effectively elevate its IP protection standards and drive high-quality economic growth. Future research could explore the application of deep learning techniques such as Mamba (Yao et al., 2024), Transformer (Wang et al., 2024), and new attention mechanisms (Huo et al., 2024) to enhance the rationale behind strategy formulation, as well as to improve model performance.